# Sample Size Selection under an Infill Asymptotic Domain

Cory W. Natoli[1], Edward D. White[1*], Beau A. Nunnally[1], Alex J. Gutman[2], and Raymond R. Hill[2]

[1]      Department of Mathematics and Statistics, Air Force Institute of Technology, Wright-Patterson Air Force Base, OH

[2]      Department of Operational Sciences, Air Force Institute of Technology, Wright-Patterson Air Force Base, OH

\*      Corresponding author

# Abstract

Experimental studies often fail to appropriately account for the number of collected samples within a fixed time interval for functional responses. Data of this nature appropriately falls under an Infill Asymptotic domain that is constrained by time and not considered infinite. Therefore, the sample size should account for this infill asymptotic domain. This paper provides general guidance on selecting an appropriate size for an experimental study for various simple linear regression models and tuning parameter values of the covariance structure used under an asymptotic domain, an Ornstein-Uhlenbeck process. Selecting an appropriate sample size is determined based on the percent of total variation that is captured at any given sample size for each parameter. Additionally, guidance on the selection of the tuning parameter is given by linking this value to the signal-to-noise ratio utilized for power calculations under design of experiments.

**Keywords:** Sample size, correlated observations, simple regression, Ornstein-Uhlenbeck process

# Sample Size Selection under an Infill Asymptotic Domain

## 1. INTRODUCTION

One of the most fundamental aspects of any statistical analysis relates to sample size determination. Design of Experiments (DOE) is often utilized to ascertain the requisite sample size to achieve a desired statistical power. In general, DOE is focused on creating efficient test plans by sizing designs based on one or more observable response variables. This response is usually the average or standard deviation of the measured characteristic (Montgomery 2017). However, many responses in practice are functions themselves, such as a time series, rather than a simple singular value. These data are known in the literature as functional data (Ramsay and Silverman 2005; Ferraty and Vieu 2006). DOE concentrates on determining the number of samples that should be taken to demonstrate that previously developed theories and understanding from observations are correct (Montgomery 2017). Often overlooked is how often within a time interval samples should be taken for these functional data.

Generally, the time interval is considered infinite, however, we know this is not likely true when conducting an experimental process. Under this scenario, the time domain, *t*, will be bounded as there will be a fixed time window from the start time to the stop time. Therefore, functional data of this nature falls under an Infill Asymptotic (IA) domain (Cressie 1993). Infinitely sampling results in variances of consistent estimators approaching zero, but variances of inconsistent estimators do not approach zero. Natoli et al. (2021) demonstrate that under an IA domain, the parameter estimators for a linear regression model are inconsistent. Further, they note a pattern of greater diminishing return regarding decreasing estimator variance as sample size increases.

That is, simply collecting more data is not efficient in either cost or time.

This paper investigates an appropriate sample size under an IA domain for either a simple linear model or a model consisting of piecewise simple linear functions. Section 2 provides a brief background on IA, the adopted Ornstein-Uhlenbeck (OU) process, and the previous work on estimator consistency and variances. In Section 3, a proposed sample size is suggested based on a ratio of a tuning parameter in the OU process, $\lambda$, and sample size, $n$. Section 4 provides details on the tuning parameter, $\lambda$, and reasonable values for this parameter under experimental studies.

## 2. BACKGROUND

This paper furthers results presented in Natoli et al. (2021) and uses the same adoption of the IA domain and covariance structure. For definition purposes, Mills (2010) remarks that the maximum distance between two observations is bounded and defines IA as:

Definition – *Suppose sampling within the domain of a process were to occur in a manner which spreads the samples as far apart as possible. If:*

$$\lim_{n \to \infty} \max_{i,j \in \{1,2,\ldots,n\}} |t_i - t_j| = C$$

where $C \in \mathcal{R}^+$ is a finite constant, $i, j \in Z^+$ are indices which order the samples, and $t_i, t_j \in \mathcal{R}^+ \cup \{0\}$ are time points corresponding to the $i^{th}$ and $j^{th}$ indices, respectively, then the domain of the process is an Infill Asymptotics (IA) domain.

Natoli et al. (2021) defines an OU process as a Gaussian process satisfying: $E(\varepsilon_{t_i}) = 0$ and $Cov(\varepsilon_{t_i}, \varepsilon_{t_j}) = \sigma^2 e^{-\lambda|t_i - t_j|}$, where $\sigma^2 > 0$ is the overall variance, $\lambda > 0$ is a tuning parameter, and $\varepsilon_{t_i}, \varepsilon_{t_j}$ are random observations in time. Additionally, $\lambda$ is assumed to be an unknown, but fixed random variable and $\sigma^2$ is equal to 1. The selection of $\lambda$, or the mean reversion parameter (Vasicek 1977, Schöbel and Zhu 1999) is critical for determining an appropriate sample size under the IA domain. This $\lambda$ must be selected a priori to adopting an appropriately sized design and is comparable to a signal-to-noise ratio (SNR) sometimes seen in DOE. Section 4 further explores how to determine a $\lambda$ value and its comparison to a SNR.

An evenly spaced sampling method is assumed for collecting observations over a fixed interval [0, T] with the first observation taken at $t_1 = 0$ and the final observation taken at $t_n = T$. Furthermore, this interval is assumed scaled to the fixed interval [0,1]. Under these assumptions, $E(\varepsilon_{t_i}) = 0$ $and$ $Cov(\varepsilon_{t_i}, \varepsilon_{t_j}) = e^{-\lambda|i-j|/(n-1)}$, with $\lambda > 0$ and $i, j \in [1, n]$. This can be written in the form $Cov(\varepsilon_{t_i}, \varepsilon_{t_j}) = \rho^{|i-j|}$ where $\rho = e^{-\lambda/(n-1)}$.

Natoli et al. (2021) provides the variance and asymptotic variance of model parameters in a simple linear regression model under three different scenarios: an intercept only model, a slope only model passing through the origin, and a model that contains both an intercept and slope, thereby allowing the *y*-intercept not to be zero. The summarized results from White (2001) and Natoli et al. (2021) are presented in Table 1.

## 3. SAMPLE SIZE CALCULATION

A ratio of the actual variance to the asymptotic variance was created to determine sample size. That is, for various values of $n$ and $\lambda$, both the actual and asymptotic variances were calculated and the ratio was created such that $Var\ Ratio = \frac{Asymptotic\ Variance}{Actual\ Variance}$. While the concept resembles that of $R^2$ value in that it describes the percent of variation, here it is describes the ratio of sample variance to asymptotic variance captured given a specific sample size for a specific $\lambda$ value, where any value less than 1 implies the sample variance will be larger. We varied values of $n$ from 3-50 and $\lambda$ from 0.001-150 to ensure sampling thoroughly throughout a potential design region.

**Table 1.** Variance and Asymptotic variance for model parameters

| Model/Parameter | Variance | Limiting Variance |
|---|---|---|
| Intercept Only $\widehat{\beta_0}$ | $\dfrac{1+\rho}{n(1-\rho)+2\rho}$ | $\dfrac{2}{2+\lambda}$ |
| Slope Only $\widehat{\beta_1}$ | $\dfrac{6(1-\rho^2)(n-1)}{2n^2(1-2\rho+\rho^2)+n(8\rho-1)+\rho^2(6-7n)}$ | $\dfrac{2}{\dfrac{1}{\lambda}+1+\dfrac{\lambda}{3}}$ |
| Mean and Slope $\widehat{\beta_0}$ | $\dfrac{2(1+\rho)[2n^2(1-2\rho+\rho^2)+n(8\rho-1-7\rho^2)+6\rho^2]}{[n(1-\rho)+2\rho][n^2(1-2\rho+\rho^2)+n(4\rho+1-5\rho^2)+6\rho(1+\rho)]}$ | $\dfrac{8(\lambda^2+3\lambda+3)}{(\lambda+2)(\lambda^2+6\lambda+12)}$ |
| Mean and Slope $\widehat{\beta_1}$ | $\dfrac{12(1-\rho^2)(n-1)}{n^2(1+\rho^2-2\rho)+n(1+4\rho-5\rho^2)+6\rho(1+\rho)}$ | $\dfrac{4}{\dfrac{2}{\lambda}+1+\dfrac{\lambda}{6}}$ |
| Mean and Slope $Cov(\widehat{\beta_0},\widehat{\beta_1})$ | $-\dfrac{6(1-\rho^2)(n-1)}{n^2(1+\rho^2-2\rho)+n(1+4\rho-5\rho^2)+6\rho(1+\rho)}$ | $\dfrac{-2}{\dfrac{\lambda}{6}+1+\dfrac{2}{\lambda}}$ |

The values of $n$ were chosen as Natoli et al. (2021) showed 50 was a sufficient number of samples to observe a point of diminishing return. Additionally, it was shown that the maximum $\lambda$ value needed is 148.9; we consequently rounded to 150. To sample the space thoroughly, a Latin Hypercube design was used with these two factors as continuous variables. An appropriate sample size was determined by observing the scatterplot matrix and determining at what sample size a homogenous arrangement of sample points existed within the selected sample region. The design selected has a maximum minimum distance of 0.091 and a discrepancy of 0. The value for $n$ was then rounded to the nearest whole number as only whole numbers are possible for the sample size and the value for $\lambda$ was rounded to the second decimal to avoid infinite strings in calculation. Figure 1 shows this scatterplot matrix.

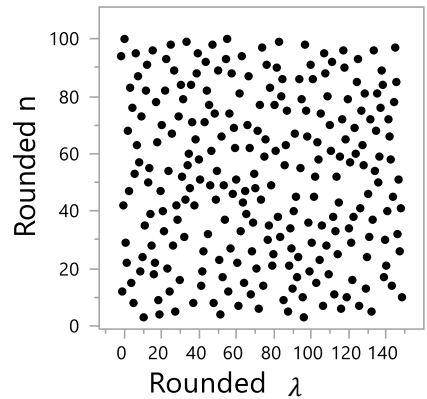

**Figure 1.** Scatterplot Matrix of Design Space

A variable for the ratio of $\lambda$ and $n$ was created. The goal is to find a consistent ratio of $\lambda$ to $n$ at which a certain amount of variation is achieved. For the tables in this section, we addressed the percent of variation for 75%, 80%, 90%, and 95%, but this value could be adjusted depending on the fidelity desired by the practitioner. A ratio of greater than or equal to 70%, would likely be of most interest for practitioners utilizing this technique. For the following results, the models fitting the percent variation versus the ratio of $\lambda$ to $n$ were constructed using only the data sampled at which the percent variation was at least 0.7 and the functions are only appropriate in the 70% to 100% range. As an illustrative example, the target variance ratio of 90% is used throughout this section with further results presented later.

The first model presented is the intercept only model. Figure 2 shows the plot for the variance ratio against the ratio of $\lambda$ to $n$ from 0 to 100%. Figure 3 shows the plot for the restricted variance ratio against the ratio of $\lambda$ to $n$ with the cubic fit overlaid on the plot.

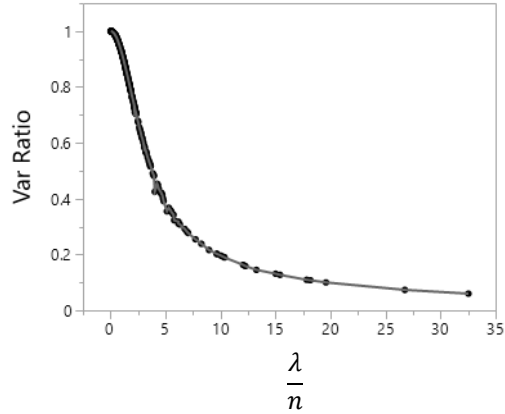

**Figure 2.** Variance ratio versus $\frac{\lambda}{n}$ for the intercept only model

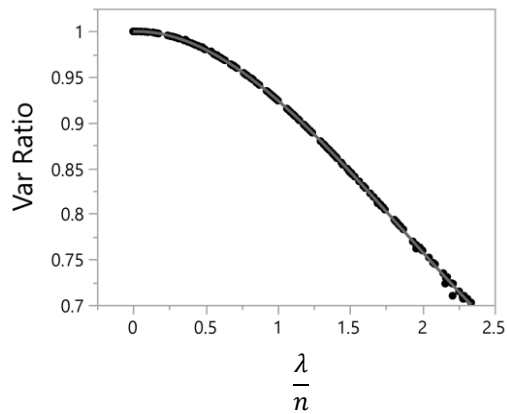

**Figure 3.** Variance ratio model from 0.7 to 1 for the intercept only model

The cubic formula used is:

$$Var\ Ratio = 1.068 - 0.143\frac{\lambda}{n} - 0.042\left(\frac{\lambda}{n} - 1.062\right)^2 + 0.017\left(\frac{\lambda}{n} - 1.062\right)^3$$

The model is an excellent fit to the data with an $R^2_{adj}$ value of 0.999. This equation crosses the Var Ratio of 0.9 at a $\frac{\lambda}{n}$ value of 1.168, which means a $\frac{\lambda}{n}$ less than 1.168 will account for at least 90% of the possible variation. However, the variance of the estimated model must be included, so the $\frac{\lambda}{n}$ ratios for the 95% confidence interval about the target variance ratios were calculated. To calculate the confidence interval, a normal approximation was used such that the confidence interval is equal to $x \pm 1.96 * RMSE$ (Root-Mean-Square Error) where $x$ is the value of $\frac{\lambda}{n}$ at a variance ratio of 0.9. For the mean only model, this is equal to $1.168 \pm 1.96 * 0.0012$. Subsequent confidence intervals are also constructed in this manner. Thus, the $\frac{\lambda}{n}$ 95% approximate confidence interval is [1.166, 1.170]. It is of interest to determine if this trend holds for the other two models.

The next model observed is the slope only model. Figure 4 shows the plot for the variance ratio against the ratio of $\lambda$ to $n$. Figure 5 shows the plot for the restricted variance ratio against the ratio of $\lambda$ to $n$ with the cubic fit overlaid on the plot.

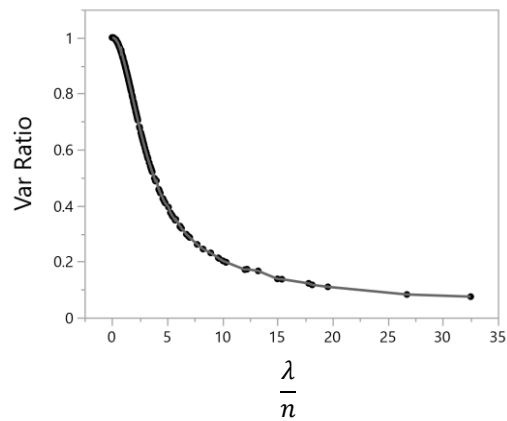

**Figure 4.** Variance ratio versus $\frac{\lambda}{n}$ for the slope only model

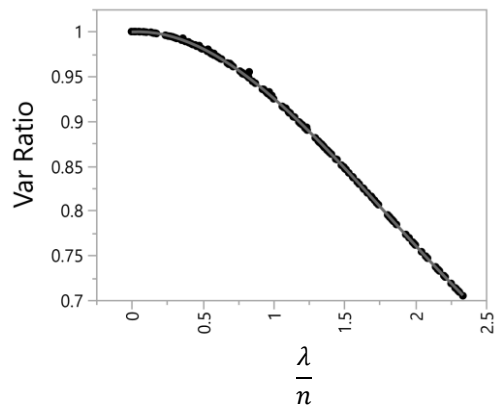

**Figure 5.** Variance ratio model from 0.7 to 1 for the slope only model

The cubic formula used is:

$$Var\ Ratio = 1.069 - 0.143\frac{\lambda}{n} - 0.041\left(\frac{\lambda}{n} - 1.062\right)^2 + 0.018\left(\frac{\lambda}{n} - 1.062\right)^3$$

The model is an excellent fit to the data with an $R^2_{adj}$ value of 0.999. This equation crosses the Var Ratio of 0.9 at a $\frac{\lambda}{n}$ value of 1.177, which means a $\frac{\lambda}{n}$ less than 1.177 will account for at least 90% of the possible variation. Again, the 95% confidence interval was used to create a range of values for $\frac{\lambda}{n}$ which is equal to [1.176,1.178].

The next model observed is a model with both an intercept and slope. First, we will look at the intercept for this model. Figure 6 shows the plot for the variance ratio of the intercept against the ratio of $\lambda$ to $n$. Figure 7 shows the plot for the restricted variance ratio of the intercept against the ratio of $\lambda$ to $n$ with the cubic fit overlaid on the plot.

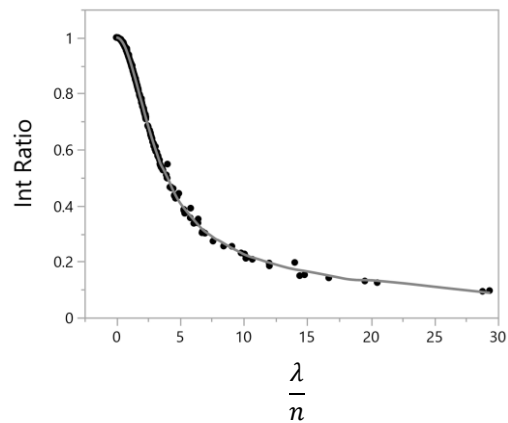

**Figure 6.** Variance ratio versus $\frac{\lambda}{n}$ for the intercept in the full model

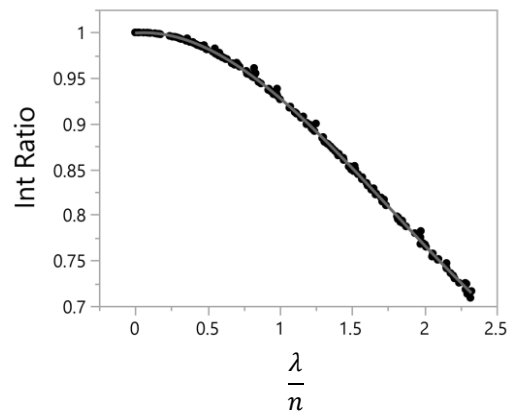

**Figure 7.** Variance ratio model from 0.7 to 1 for the intercept in the full model

The cubic formula used is:

$$Int\ Var\ Ratio = 1.069 - 0.140\frac{\lambda}{n} - 0.041\left(\frac{\lambda}{n} - 1.062\right)^2 + 0.019\left(\frac{\lambda}{n} - 1.062\right)^3$$

The model is an excellent fit to the data with an $R^2_{adj}$ value of 0.999. This line crosses the Var Ratio of 0.9 at a $\frac{\lambda}{n}$ value of 1.195, which means a $\frac{\lambda}{n}$ less than 1.195 will account for at least 90% of the possible variation. The 95% confidence interval was used to create a range of values for $\frac{\lambda}{n}$ which is equal to [1.192,1.199].

Next, we will look at the slope for this model. Figure 8 shows the plot for the variance ratio of the slope against the ratio of $\lambda$ to $n$. Figure 9 shows the plot for the restricted variance ratio of the slope against the ratio of $\lambda$ to $n$ with the cubic fit overlaid on the plot.

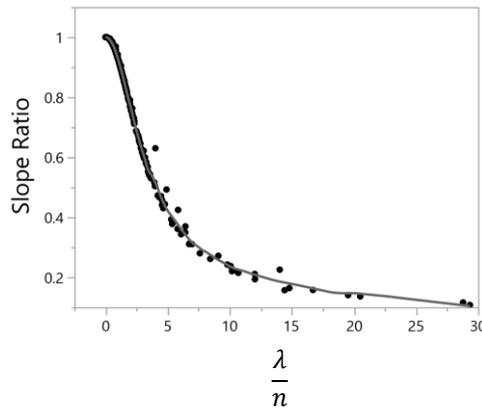

**Figure 8.** Variance ratio versus $\frac{\lambda}{n}$ for the slope

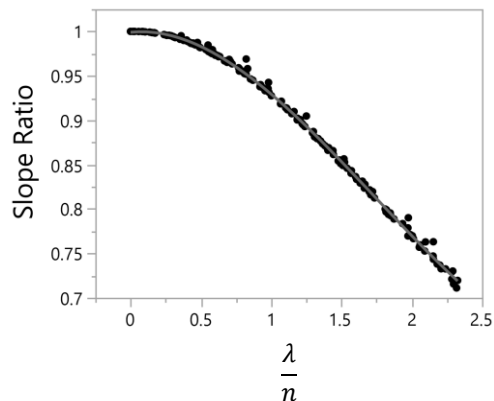

**Figure 9.** Variance ratio model from 0.7 to 1 for the slope in the full model

The cubic formula used is:

$$Var\ Ratio = 1.069 - 0.140\frac{\lambda}{n} - 0.041\left(\frac{\lambda}{n} - 1.062\right)^2 + 0.020\left(\frac{\lambda}{n} - 1.062\right)^3$$

The model is an excellent fit to the data with an $R^2_{adj}$ value of 0.998. This line crosses the Var Ratio of 0.9 at a $\frac{\lambda}{n}$ value of 1.205, which means a $\frac{\lambda}{n}$ less than 1.205 will account for at least 90% of the possible variation. Again, this matches with what was seen in the intercept and slope only models. The 95% confidence interval was used to create a range of values for $\frac{\lambda}{n}$ and equals [1.200, 1.210].

Finally, we look at the covariance for this model. Figure 10 shows the plot for the covariance ratio against the ratio of $\lambda$ to $n$. Figure 11 shows the plot for the restricted covariance ratio against the ratio of $\lambda$ to $n$ with the cubic fit overlaid on the plot.

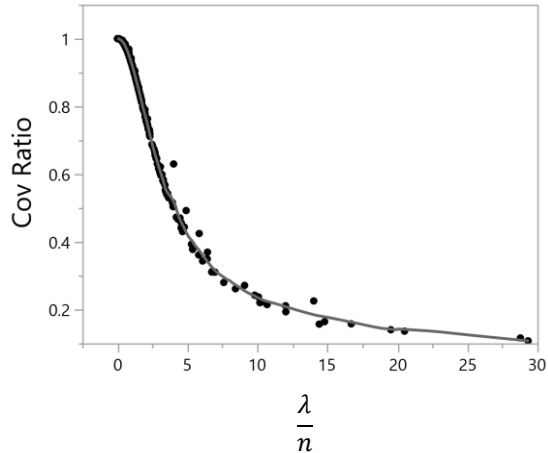

Figure 10. Covariance ratio versus $\frac{\lambda}{n}$

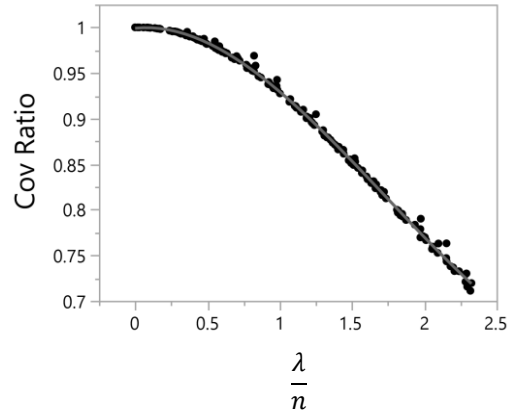

Figure11. Covariance ratio model from 0.7 to 1

The cubic formula used is:

$$Var\ Ratio = 1.069 - 0.140\frac{\lambda}{n} - 0.041\left(\frac{\lambda}{n} - 1.062\right)^2 + 0.020\left(\frac{\lambda}{n} - 1.062\right)^3$$

The model is an excellent fit to the data with an $R^2_{adj}$ value of 0.998. This line crosses the Covariance Ratio of 0.9 at a $\frac{\lambda}{n}$ value of 1.205, which means a $\frac{\lambda}{n}$ less than 1.205 will account for at least 90% of the possible covariance. A 95% confidence interval was used to create a range of values for $\frac{\lambda}{n}$ and equals [1.200, 1.210].

Since the shape and values for each of the previous models are so similar, it would be useful if a single value could be used independently of the model type being fit. That is, this agnostic model can be used if the form of the data is unknown. To determine if a single $\frac{\lambda}{n}$ value could be used for all of the models, all of the data was combined. That is, the variance ratio of mean in the mean only model, the slope in the slope only model, and the slope, mean, and covariance in the full model were all used as the variance ratio and plotted against the $\frac{\lambda}{n}$ ratio. Again, a cubic model was fit to this data. This model can be seen in Figure 12.

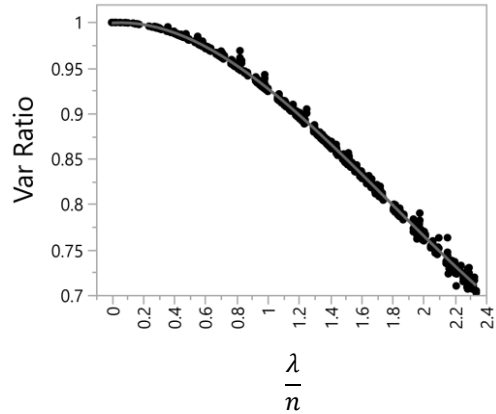

Figure 12. Variance ratio model from 0.7 to 1 for the combined data

The cubic formula used is:

$$Var\ Ratio = 1.069 - 0.141\frac{\lambda}{n} - 0.041\left(\frac{\lambda}{n} - 1.062\right)^2 + 0.019\left(\frac{\lambda}{n} - 1.062\right)^3$$

The model is a good fit to the combined data with an $R^2_{adj}$ value of 0.998 indicating this is an appropriate model for the data. This line crosses the Variance Ratio of 0.9 at a $\frac{\lambda}{n}$ value of 1.190, which means a $\frac{\lambda}{n}$ less than 1.190 will account for at least 90% of the possible covariance. A 95% confidence interval was used to create a range of values for $\frac{\lambda}{n}$ and equals [1.188, 1.193].

The results show that for any of the models presented here and for any given $\lambda$, an appropriate sample size can be calculated. Given a specific model, the sample size can be accounted for using the model specific variance ratios, or a more generalized model can be utilized. The results of each model are presented in Tables 2-5.

Table 2. Confidence intervals for 75% Variance Ratio

| Model/Parameter | 95% CI Lower Bound | $\frac{\lambda}{n}$ at 75% Var Ratio | 95% CI Upper Bound |
|---|---|---|---|
| Intercept Only $\widehat{\beta_0}$ | 2.045 | 2.047 | 2.049 |
| Slope Only $\widehat{\beta_1}$ | 2.065 | 2.066 | 2.068 |
| Mean and Slope $\widehat{\beta_0}$ | 2.096 | 2.101 | 2.105 |
| Mean and Slope $\widehat{\beta_1}$ | 2.116 | 2.123 | 2.131 |
| Mean and Slope $Cov(\widehat{\beta_0},\widehat{\beta_1})$ | 2.116 | 2.123 | 2.131 |
| Agnostic Model | 2.088 | 2.091 | 2.094 |

Table 3. Confidence intervals for 80% Variance Ratio

| Model/Parameter | 95% CI Lower Bound | $\frac{\lambda}{n}$ at 80% Var Ratio | 95% CI Upper Bound |
|---|---|---|---|
| Intercept Only $\widehat{\beta_0}$ | 1.762 | 1.764 | 1.766 |
| Slope Only $\widehat{\beta_1}$ | 1.776 | 1.778 | 1.779 |
| Mean and Slope $\widehat{\beta_0}$ | 1.800 | 1.804 | 1.807 |
| Mean and Slope $\widehat{\beta_1}$ | 1.813 | 1.819 | 1.824 |
| Mean and Slope $Cov(\widehat{\beta_0},\widehat{\beta_1})$ | 1.813 | 1.819 | 1.824 |
| Agnostic Model | 1.794 | 1.796 | 1.799 |

Table 4. Confidence intervals for 90% Variance Ratio

| Model/Parameter | 95% CI Lower Bound | $\frac{\lambda}{n}$ at 90% Var Ratio | 95% CI Upper Bound |
|---|---|---|---|
| Intercept Only $\widehat{\beta_0}$ | 1.166 | 1.168 | 1.170 |
| Slope Only $\widehat{\beta_1}$ | 1.176 | 1.177 | 1.178 |
| Mean and Slope $\widehat{\beta_0}$ | 1.192 | 1.195 | 1.199 |
| Mean and Slope $\widehat{\beta_1}$ | 1.200 | 1.205 | 1.210 |
| Mean and Slope $Cov(\widehat{\beta_0}, \widehat{\beta_1})$ | 1.200 | 1.205 | 1.210 |
| Agnostic Model | 1.188 | 1.190 | 1.193 |

Table 5. Confidence intervals for 95% Variance Ratio

| Model/Parameter | 95% CI Lower Bound | $\frac{\lambda}{n}$ at 95% Var Ratio | 95% CI Upper Bound |
|---|---|---|---|
| Intercept Only $\widehat{\beta_0}$ | 0.799 | 0.802 | 0.804 |
| Slope Only $\widehat{\beta_1}$ | 0.810 | 0.812 | 0.813 |
| Mean and Slope $\widehat{\beta_0}$ | 0.823 | 0.828 | 0.832 |
| Mean and Slope $\widehat{\beta_1}$ | 0.831 | 0.838 | 0.844 |
| Mean and Slope $Cov(\widehat{\beta_0}, \widehat{\beta_1})$ | 0.831 | 0.838 | 0.844 |
| Agnostic Model | 0.820 | 0.823 | 0.826 |

The results make it clear that as the percent of variation being accounted for increases, the $\frac{\lambda}{n}$ ratio required decreases. If $\lambda$ were held constant, a decrease in $\frac{\lambda}{n}$ would require an increase in sample size. Therefore, the sample size required is increasing as the percent of variation being accounted for increases, which is expected as it takes more samples to account for more of the total variation. These results provide a useful estimate of the $\lambda/n$ ratio needed, while the next section focuses on selecting an appropriate $\lambda$ value for an experiment.

# 4. SELECTING $\lambda$

Selecting an appropriate $\lambda$ is an important, yet difficult, step for determining the sample size. Determining the value for this tuning parameter is still an active area of research (Hu and Nualart 2010; Sánchez and Palacio 2013; Thomas, et al. 2015). $\lambda$ is considered the mean reversion parameter (Vasicek 1977; Schöbel and Zhu 1999). The larger the $\lambda$, the slower the function converges to the true mean. The choice for $\lambda$ under this scenario is in line with Cressler et al. (2015). The purpose of the tuning parameter under this scenario is the mean reversion of an evolutionary process to a true model. Cressler et al. state that this range represents a broad spectrum of scenarios, from ones in which the data contain very little information about the evolutionary process to ones where recovery of the true model is nearly guaranteed. While a range of values is useful for practical application, an even more useful property noted by Cressler et al. is the relationship to the signal-to-noise ratio (SNR).

SNR is a metric that is used to size experiments under a typical DOE scenario with the "signal" being the difference in the response the tester wishes to detect and "noise" being the variation in the system. More important than the SNR values themselves is the value of the ratio. The ratio determines the signal the practitioner wishes to detect as a multiple of the standard deviation. For example, a SNR of 2 would indicate that the practitioner wishes to detect a difference of two standard deviations. The SNR is also referred to as Cohen's d (Cohen 1988). Cohen suggested that under his psychological research values of d of 0.2, 0.4 or 0.8 should be considered as "small", "medium" and "large" effect sizes, respectively (Cohen 1988). However, the definition of "small", "medium", and "large" can vary by areas of study. Larger effects than these presented by Cohen can be achieved (Harris and Rice 2003; 2005). Thus, when designing a test, one must recognize the difference between a practical and statistical significance (Montgomery 2017). The difference depends on the field of study. Within the Department of Defense, the Institute for Defense Analysis (IDA) conducted studies on SNRs in operational testing. From this, IDA concluded that 56% of SNRs lie between 0.5 and 2 (Avery 2014), with no SNRs that exceeded 7. The potential values of SNRs are highly dependent on the field of study, and will require the assistance of subject matter experts. There are many ways to determine a reasonable SNR for a test, such as using previously designed experiments, talking to subject matter experts, or a conducting a pilot study to estimate system variation.

The reason that linking SNR to $\lambda$ is critical is that, like the SNR, the $\lambda$ value must be determined a priori to appropriately size the test. Cressler et al. notes that SNR is equal to $\sqrt{\lambda}\sigma^2$. However, the assumption in this paper is that $\sigma^2$ is equal to 1, so the SNR is simply equal to $\sqrt{\lambda}$. Therefore, $\lambda = (SNR)^2$. Under this scenario, a $\lambda$ is determined by specifying a desired SNR prior to testing. Continuing with the example of a SNR of 2, $\lambda$ would equal 4.

As an example, assume that we are seeking to meet a 90% variance ratio in an experiment. If the SNR desired for the experiment is 2, which implies $\lambda = 4$, then the sample size can be calculated for the experiment. If the form of the model needed to represent the data is unknown, use the agnostic model. Thus, to have at least 90% of the variance explained, we have the inequality $\frac{4}{n} < 1.190$, so $n > 3.36$. Thus, a sample size of 4 is required to account for at least 90% of the variation. This scales depending on $\sigma^2$. Continuing with this example, if $\sigma^2$ were actually 0.5 instead of 1, the value of $\lambda$ would be calculated such that $\sqrt{\lambda} * 0.5 = 2$ and $\lambda = 16$ yielding $n > 19.04$.

# 5. CONCLUSION

An IA domain presents a unique problem in assessing sample size as the parameters of a linear regression model are inconsistent. However, by using the variance and asymptotic variances of the parameters, an appropriate sample size for experimental studies can be determined under an IA domain. A rule of thumb for accounting for 90% of the variation is to ensure that the $\frac{\lambda}{n}$ ratio is less than 1. The amount of variation accounted for can be adjusted as necessary and a $\frac{\lambda}{n}$ ratio can be determined using the cubic formulas presented. Moreover, a $\lambda$ value can be calculated a priori through the use of a SNR. This process mirrors a classical DOE process to assess a design. The results presented in this paper allow a practitioner to answer the question "How many samples should I take?" under an IA domain. This provides a useful tool when assessing any temporal data or single dimension functional data with a limited domain.